\documentclass[journal]{IEEEtran}
\usepackage[pdftex]{graphicx}
\usepackage{amsmath,mathtools,algorithm,cite}
\allowdisplaybreaks[4]
\usepackage{amsfonts,amssymb,amsthm,algpseudocode}
\usepackage{bm}
\title{
A Multi-Target Track-Before-Detect Particle Filter
Using Superpositional Data in Non-Gaussian Noise
} 
\author{Nobutaka Ito \IEEEmembership{Member, IEEE} and Simon Godsill \IEEEmembership{Senior Member, IEEE}
\thanks{Nobutaka Ito and Simon Godsill are with
Department of Engineering, University of Cambridge, United Kingdom (e-mail: sjg@eng.cam.ac.uk). }}
\begin{document}
\maketitle

\begin{abstract}
This paper proposes a novel particle filter for tracking time-varying states of multiple targets jointly from \textbf{\textit{superpositional}} data, which depend on the sum of contributions of all targets. Many conventional tracking methods rely on preprocessing for detection (\textbf{\textit{e.g.}}, thresholding), which severely limits tracking performance at a low signal-to-noise ratio (SNR). In contrast, the proposed method can operate directly on raw sensor signals without requiring such preprocessing. Though there also exist methods applicable to raw sensor signals called \textbf{\textit{track-before-detect}}, the proposed method has significant advantages over them. First, it is general without any restrictions on observation/process noise statistics (\textbf{\textit{e.g.}}, Gaussian) or the functional form of each target's contribution to the sensors (\textbf{\textit{e.g.}}, linear, separable, binary). Especially, it includes Salmond \textbf{\textit{et al.}}'s track-before-detect particle filter for a single target as a particular example up to some implementation details. Second, it can track an unknown, time-varying number of targets without knowing their initial states owing to a target birth/death model. We present a simulation example of radio-frequency tomography, where it significantly outperformed Nannuru \textbf{\textit{et al.}}'s state-of-the-art method based on random finite sets in terms of the optimal subpattern assignment (OSPA) metric.
\end{abstract}
\begin{IEEEkeywords}
Multi-target tracking (MTT),
track-before-detect,
particle filter,
superpositional sensor signals.
\end{IEEEkeywords}
\section{Introduction}
Multi-target tracking (MTT) aims to estimate time-varying states of multiple targets jointly from available observations,
where these states typically include kinematic states ({\it e.g.,} the position/velocity/acceleration).
MTT constitutes one of the most active areas in statistical signal processing, and 
 myriads of methods have been proposed.
These methods
encompass multiple hypothesis tracking (MHT)~\cite{Reid1979},
joint probabilistic data association (JPDA)~\cite{Fortmann1983},
 probabilistic multihypothesis tracking (PMHT)~\cite{Streit1994}, random finite sets (RFS)~\cite{Mahler2003,Mahler2007,Vo2005,Vo2006},
multi-target particle filters~\cite{Hue2002,Vermaak2005},
and sequential Markov chain Monte Carlo (MCMC)~\cite{Septier2009}.

The conventional approach to MTT is based on
a two-step procedure, consisting of 
detection and tracking steps.
In the detection step,
sensor signals are preprocessed for detection by, {\it e.g.,} thresholding, 
so that each detection corresponds to a single target or a clutter.
These detections are fed into the subsequent tracking step.
In such a procedure, tracking performance heavily depends on detection performance.
Unfortunately, the latter degrades severely at a low signal-to-noise ratio (SNR), and thus, so does the former.
Note that, 
in this approach, information contained in observed time series is not being fully exploited for detection,
because it is performed based solely on observations at the current time step 
without reference to past time steps.

At a low SNR,
it is significantly advantageous to operate directly on
raw sensor signals 
for joint detection and tracking,
which is
known in the literature
{\it track-before-detect}.
Compared to the two-step procedure, 
this approach can exploit information from
not only the current 
but also past time steps in performing detection,
leading to better noise robustness.

Salmond {\it et al.}~\cite{Salmond2001} proposed
a track-before-detect method for 
at most one target
in the framework of recursive Bayesian estimation.
Since then several authors
have considered multi-target extensions of track-before-detect.
Kreucher {\it et al.}
 and Vo {\it et al.}
focused on a restricted class of sensor signals,
where
targets contribute in a binary~\cite{Kreucher2005} or  a disjoint~\cite{Vo2010} manner.
Mahler~\cite{Mahler2009,Mahler2014} 
derived a RFS-based filter
for multi-target track-before-detect using
{\it superpositional} sensor signals, 
which depend on the sum of general nonlinear  target contributions. 
This method is called a superpositional
cardinalized probability hypothesis density ($\Sigma$-CPHD) filter~\cite{Mahler2014}.
Moreover,
Nannuru {\it et al.}~\cite{Nannuru2013} developed a tractable approximate implementation of the
$\Sigma$-CPHD filter based on the particle filter, but it is limited to additive Gaussian observation noise.
Several authors~\cite{Boers2003,Lepoutre2013,Cuillery2019} 
considered
superpositional sensor signals in the presence of 
unknown target amplitudes,
but again this is limited to additive Gaussian observation noise.
Orton {\it et al.}~\cite{Orton2002} proposed a
particle filter for multi-target track-before-detect
using superpositional sensor signals
on the assumption that the number of targets is given.

Here we propose a novel method
for
multi-target track-before-detect using superpositional sensor signals.
It can
1) operate on superpositional sensor signals
involving nonlinear target contributions of general functional forms
and
non-Gaussian observation/process noise to
 2) track
an unknown, time-varying number of targets 3) without knowing their initial states 4) in an online manner.
The proposed method is a multi-target extension of
 Salmond {\it et al.}'s single-target track-before-detect particle filter~\cite{Salmond2001} based on Septier {\it et al.}'s state modeling~\cite{Septier2009} with a birth/death process.
This state modeling enables the proposed method to deal with an unknown, time-varying number of targets.


\section{Proposed Particle Filter for\\ Multi-Target Track-Before-Detect}
This section describes the proposed 
particle filter for multi-target track-before-detect.
It is based on a state-space model
with time index $t$, unknown states $\widetilde{\mathbf{x}}_t\,(t=0,1,2,\ldots)$, a given initial distribution $p(\widetilde{\mathbf{x}}_0)$,
a given transition distribution $p(\widetilde{\mathbf{x}}_t\mid \widetilde{\mathbf{x}}_{t-1})$, given observations $\mathbf{z}_t\,(t=1,2,\ldots)$,
and a given observation distribution $p(\mathbf{z}_t\mid \widetilde{\mathbf{x}}_t)$. 

\subsection{Modeling Multi-Target States}
Our states and their modeling follow Septier {\it et al.}~\cite{Septier2009}.
In MTT, a target may enter/leave the region observed by sensors or start/cease to emit a signal at any time, referred to as target birth/death. Consequently,
 the number of active targets is unknown and time-varying in general.
To deal with such a general setting, we prepare $n_{max}$ target models, which may be active or inactive at each time $t$, with $n_{max}$ being the maximum possible number (given) of simultaneously active targets. 
Each target model $j\in\{1,\ldots,n_{max}\}$ has a discrete state $a_{jt}\in\{0,1\}$
indicating whether it is active ($a_{jt}=1$) or not ($a_{jt}=0$) at each time $t$.
It also has continuous states
 $\mathbf{x}_{jt}\in\mathbb{R}^{n_x}$, which usually include kinematic states and possibly target signal amplitude~\cite{Rollason2001,Rollason2018} or
any other physical quantities.
Our states $\widetilde{\mathbf{x}}_t$ consist of 
$
\mathbf{x}_t\coloneqq (\mathbf{x}_{1t},\ldots,\mathbf{x}_{n_{max},t})\in\mathbb{R}^{n_{max}n_x}$ and
$\mathbf{a}_t\coloneqq ({a}_{1t},\ldots,{a}_{n_{max},t})\in\{0,1\}^{n_{max}}$.

We assume that the transition distribution factorizes as 
\begin{align}
&p(\mathbf{x}_t,\mathbf{a}_t\mid\mathbf{x}_{t-1},\mathbf{a}_{t-1})\notag\\
&=\prod_{j=1}^{n_{max}}
p(\mathbf{x}_{jt},a_{jt}\mid\mathbf{x}_{j,t-1},a_{j,t-1})\label{eq:decompsingle}\\
&=\prod_{j=1}^{n_{max}}
\underbrace{P(a_{jt}\mid a_{j,t-1})}_{\text{transition of discrete states}}
\underbrace{p(\mathbf{x}_{jt}\mid\mathbf{x}_{j,t-1},a_{jt},a_{j,t-1})}_{\text{transition of continuous states}}.\label{eq:fac}
\end{align}
The factor
$P(a_{jt}\mid a_{j,t-1})$
 is modeled by a transition probability matrix
\begin{align}
\begin{pmatrix}
1-\pi_b&\pi_b\\
\pi_d&1-\pi_d
\end{pmatrix},\label{eq:TPM}
\end{align}
where $\pi_b$ and $\pi_d$ are given birth and death probabilities.
The factor $p(\mathbf{x}_{jt}\mid\mathbf{x}_{j,t-1},a_{jt},a_{j,t-1})$ is modeled as
\begin{align}
&p(\mathbf{x}_{jt}\mid\mathbf{x}_{j,t-1},a_{jt},a_{j,t-1})\notag\\
&=\begin{cases}
p_s(\mathbf{x}_{jt}\mid\mathbf{x}_{j,t-1}),&\text{if }(a_{jt}, a_{j,t-1})=(1,1)\\
p_b(\mathbf{x}_{jt}),&\text{if }(a_{jt}, a_{j,t-1})=(1,0)\\
p_d(\mathbf{x}_{jt}),&\text{if }a_{jt}=0.
\end{cases}\label{eq:transcont}
\end{align}
Here, $p_s$ and $p_b$ are given densities corresponding to target survival and birth, respectively, where $p_s$ 
 may be non-Gaussian and involve nonlinearity. 
As we will see later, 
$p_d$ is actually not used in our particle filter at all,
and therefore does not need to be specified.

\subsection{Modeling Observations}
In our track-before-detect setting, observations $\mathbf{z}_t\in\mathbb{C}^{n_z}$ consist in raw sensor signals,
where $n_z$ denotes the number of sensor signals.
Let us first consider the case of additive noise for simplicity, where $\mathbf{z}_t$ is modeled as
$
\mathbf{z}_t=\sum_{j=1}^{n_{max}}
a_{jt}
h(\mathbf{x}_{jt})+\mathbf{v}_t=\sum_{j:\,a_{jt}=1}
h(\mathbf{x}_{jt})+\mathbf{v}_t.
$
Here, 
$h(\mathbf{x}_{jt})$ is the target signal from target $j$ with $h: \mathbb{R}^{n_x}\rightarrow\mathbb{C}^{n_z}$ being a given, possibly nonlinear function, and
$\sum_{j=1}^{n_{max}}
a_{jt}
h(\mathbf{x}_{jt})$ 
the sum of the target signals from all active targets.
Additive noise
$\mathbf{v}_t$ is assumed to be independent from time step to time step and have
 a given, possibly non-Gaussian distribution $p_v$.
In this case,
 the observation distribution is given by
\begin{align}
p(\mathbf{z}_t\mid\mathbf{x}_t,\mathbf{a}_t)
=p_v\Biggl(\mathbf{z}_t-\sum_{j=1}^{n_{max}}
a_{jt}
h(\mathbf{x}_{jt})\Biggr).\label{eq:Gaussian}
\end{align}
This model corresponds to Mahler~\cite{Mahler2009}, 
and
 specifically to Nannuru {\it et al.}~\cite{Nannuru2013} 
in the Gaussian case.

However, here we consider a more general observation distribution, which is such that it depends on 
$\widetilde{\mathbf{x}}_t=(\mathbf{x}_t,\mathbf{a}_t)$ only through $\sum_{j=1}^{n_{max}}
a_{jt}
h(\mathbf{x}_{jt})$. That is, we consider an observation distribution of form
\begin{align}
p(\mathbf{z}_t\mid\mathbf{x}_t,\mathbf{a}_t)
&=p_o\Biggl(\mathbf{z}_t\mid \sum_{j=1}^{n_{max}}
a_{jt}
h(\mathbf{x}_{jt})\Biggr),\label{eq:obsden}
\end{align}
where $p_o$ 
is a given distribution possibly involving nonlinearity and non-Gaussianity.

\subsection{Recursive Bayesian Estimation}
In the Bayesian framework, 
we aim to obtain a posterior distribution $p(\mathbf{x}_t,\mathbf{a}_t\mid\mathbf{z}_{1:t})$ of the states 
given all observations up to the current time $t$,
where ${1:t}$ is a shorthand notation for $1,\ldots,t$.
This can be done recursively by alternating prediction and update steps, which can be carried out for our hybrid discrete-continuous states $(\mathbf{x}_t,\mathbf{a}_t)$ as well
in a similar manner to \cite{McGinnity2000,Angelova2000,Musso2001,Boers2003}.

\begin{algorithm}
\caption{Proposed particle filter for multi-target track-before-detect.}
\label{algo}
\hspace*{\algorithmicindent}\textbf{Input:} $\{\mathbf{x}_{t-1}^{k}, \mathbf{a}_{t-1}^{k}, w_{t-1}^{k}\}_{k=1}^{n_p}, \mathbf{z}_t$\vspace{1mm}\\
\hspace*{\algorithmicindent}\textbf{Output:} $\{\mathbf{x}_{t}^{k}, \mathbf{a}_{t}^{k}, w_{t}^{k}\}_{k=1}^{n_p}$
\begin{algorithmic}[1]
\For{$k=1:n_p$}
\For{$j=1:n_{max}$}
\State Draw $\widetilde{a}_{jt}^k\sim P(a_{jt}\mid a_{j,t-1}^k)$\vspace{0.5mm}
\If{$(\widetilde{a}_{jt}^k, a_{j,t-1}^k)=(1,1)$}
\State Draw $\widetilde{\mathbf{x}}_{jt}^k\sim p_s(\mathbf{x}_{jt}\mid\mathbf{x}_{j,t-1}^k)$
\EndIf
\If{$(\widetilde{a}_{jt}^k, a_{j,t-1}^k)=(1,0)$}
\State Draw $\widetilde{\mathbf{x}}_{jt}^k\sim p_b(\mathbf{x}_{jt})$
\EndIf
\EndFor
\State $\widetilde{\mathbf{s}}_t^k\leftarrow\sum_{j:\,\widetilde{a}_{jt}^k=1}h(\widetilde{\mathbf{x}}_{jt}^k)$
\State $\widetilde{w}_{t}^k\leftarrow p_o(\mathbf{z}_t\mid \widetilde{\mathbf{s}}_t^k)w_{t-1}^k$
\EndFor
\State Resample from $\{\widetilde{w}_{t}^k\}_{k=1}^{n_p}$ to get $\{k^l\}_{l=1}^{n_p}$, where $k^l$ denotes
the index of the parent of the $l$th resampled particle 
\For{$l=1:n_p$}
\For{$j=1:n_{max}$}
\State Draw $a_{jt}^l\sim P(a_{jt}\mid a_{j,t-1}^{k^l})$\vspace{0.5mm}
\If{$(a_{jt}^{l}, a_{j,t-1}^{k^l})=(1,1)$}
\State Draw $\mathbf{x}_{jt}^l\sim p_s(\mathbf{x}_{jt}\mid\mathbf{x}_{j,t-1}^{k^l})$
\EndIf
\If{$(a_{jt}^{l}, a_{j,t-1}^{k^l})=(1,0)$}
\State Draw $\mathbf{x}_{jt}^l\sim p_b(\mathbf{x}_{jt})$
\EndIf
\EndFor
\State ${\mathbf{s}}_t^l\leftarrow\sum_{j:\,{a}_{jt}^l=1}h({\mathbf{x}}_{jt}^l)$
\State $w_{t}^l\leftarrow \displaystyle\frac{p_o(\mathbf{z}_t\mid {\mathbf{s}}_t^l)}{p_o(\mathbf{z}_t\mid \widetilde{\mathbf{s}}_t^{k^l})}$
\EndFor
\State Normalize $\{w_{t}^{l}\}_{l=1}^{n_p}$ so that $\sum_{l=1}^{n_p}w_{t}^l=1$
\end{algorithmic}
\end{algorithm}

Suppose the posterior distribution $p(\mathbf{x}_{t-1},\mathbf{a}_{t-1}\mid\mathbf{z}_{1:t-1})$ at time $t-1$ is available.
The prediction step uses the transition distribution 
to obtain a prediction distribution $p(\mathbf{x}_{t},\mathbf{a}_{t}\mid\mathbf{z}_{1:t-1})$ by the Chapman-Kolmogorov equation:
\begin{align}
&p(\mathbf{x}_t,\mathbf{a}_t\mid\mathbf{z}_{1:t-1})\label{eq:CK}\\
&=\sum_{\mathbf{a}_{t-1}}\int p(\mathbf{x}_t,\mathbf{a}_t\mid \mathbf{x}_{t-1},\mathbf{a}_{t-1})p(\mathbf{x}_{t-1},\mathbf{a}_{t-1}\mid\mathbf{z}_{1:t-1})d\mathbf{x}_{t-1}.\notag
\end{align}
Here, $\sum_{\mathbf{a}_{t-1}}$ denotes the sum over $\mathbf{a}_{t-1}\in\{0,1\}^{n_{max}}$,
and we define $p(\mathbf{x}_{0},\mathbf{a}_{0}\mid\mathbf{z}_{1:0})\coloneqq p(\mathbf{x}_{0},\mathbf{a}_{0})$ with similar notations defined analogously.
The update step combines this prediction distribution
with the observation distribution to obtain the posterior distribution $p(\mathbf{x}_t,\mathbf{a}_t\mid\mathbf{z}_{1:t})$ at time $t$ by the Bayes theorem:
\begin{align}
&p(\mathbf{x}_t,\mathbf{a}_t\mid\mathbf{z}_{1:t})=\frac{p(\mathbf{z}_t\mid\mathbf{x}_t,\mathbf{a}_t)p(\mathbf{x}_t,\mathbf{a}_t\mid\mathbf{z}_{1:t-1})}{p(\mathbf{z}_t\mid\mathbf{z}_{1:t-1})}.\label{eq:Bayes}
\end{align}
Here, the normalizing constant in the denominator writes
$
p(\mathbf{z}_t\mid\mathbf{z}_{1:t-1})=\sum_{\mathbf{a}_{t}}\int p(\mathbf{z}_t\mid\mathbf{x}_t,\mathbf{a}_t)p(\mathbf{x}_t,\mathbf{a}_t\mid\mathbf{z}_{1:t-1})d\mathbf{x}_t.
$

\subsection{Particle Filter Implementation}
The Bayesian recursion in (\ref{eq:CK}) and (\ref{eq:Bayes}) can be implemented
by using the particle filter (also known as sequential Monte Carlo)~\cite{Kitagawa1987,Gordon1993,Doucet2001,Ristic}. 
It is a versatile framework applicable to the general nonlinear, non-Gaussian state-space model, where
the posterior distribution $p(\mathbf{x}_t,\mathbf{a}_t\mid\mathbf{z}_{1:t})$ is approximated by using $n_p$ point masses (or ``particles'') as
\begin{align}
p(\mathbf{x}_t,\mathbf{a}_t\mid\mathbf{z}_{1:t})\approx \sum_{k=1}^{n_p}w^k_t\delta_{\mathbf{x}^k_t}(\mathbf{x}_t)\delta_{\mathbf{a}^k_t}(\mathbf{a}_t).
\end{align}
Here,
$\{\mathbf{x}_t^k,\mathbf{a}_t^k\}_{k=1}^{n_p}$ denotes particle locations,
$\{w^k_t\}_{k=1}^{n_p}$ probability masses located at the particle locations satisfying $\sum_{k=1}^{n_p}w^k_t=1$, 
$\delta_{\mathbf{x}^k_t}(\mathbf{x}_t)$ the Dirac delta function located at $\mathbf{x}^k_t$, and
$\delta_{\mathbf{a}^k_t}(\mathbf{a}_t)$ the Kronecker delta
\begin{align}
\delta_{\mathbf{a}^k_t}(\mathbf{a}_t)
=\begin{cases}
1,&\text{if }\mathbf{a}_t=\mathbf{a}^k_t\\
0,&\text{otherwise.}
\end{cases}
\end{align}
The particle filter recursively computes particles 
$\{\mathbf{x}_{t}^{k}, \mathbf{a}_{t}^{k}, w_{t}^{k}\}_{k=1}^{n_p}$
at each time $t$, given 
particles 
$\{\mathbf{x}_{t-1}^{k}, \mathbf{a}_{t-1}^{k}, w_{t-1}^{k}\}_{k=1}^{n_p}$
 at the previous time $t-1$ and observations $\mathbf{z}_t$.

There are several implementations of the particle filter, 
and here we use 
an auxiliary particle filter~\cite{Pitt1999} (see also \cite{Ristic}).
This implementation takes account of observations at time $t$
when
 generating particle locations at time $t$, and 
can be more effective than the simple sequential importance resampling (SIR) implementation.
Algorithm~\ref{algo} presents
a pseudocode of one iteration of the proposed auxiliary particle filter for multi-target track-before-detect.
Before applying Algorithm~\ref{algo}, we initialize
the particles by $(\mathbf{x}_0^k,\mathbf{a}_0^k)\sim p(\mathbf{x}_0,\mathbf{a}_0)$, $w_0^k=1/n_p\,(k=1,\ldots,n_p)$.
It is assumed that $h$ can be evaluated at any point, 
and so does $p_o$ up to a normalizing constant.
It is also assumed that it is possible to sample realizations from $p_s$, $p_b$, and $p(\mathbf{x}_0,\mathbf{a}_0)$.
We perform ancestral sampling based on the factorization (\ref{eq:fac}) to
sample realizations from the transition distribution
$
p({\mathbf{x}}_t,{\mathbf{a}}_t\mid
{\mathbf{x}}_{t-1},{\mathbf{a}}_{t-1})$.

\subsection{Point Estimation}
Once a particle representation of the posterior probability
is obtained, it can be used to compute various point estimates of the states. In this paper, we focus on minimum mean square error (MMSE) type estimates.
The MMSE estimate of $\mathbf{x}_{jt}$ conditional to $a_{jt}=1$ can be computed by
\begin{align}
\widehat{\mathbf{x}}^{MMSE}_{jt}\coloneqq\mathbb{E}[\mathbf{x}_{jt}\mid\mathbf{z}_{1:t},a_{jt}=1]=\frac{\sum_{k=1}^{n_p}w_t^ka_{jt}^k\mathbf{x}_{jt}^k
}{\sum_{k=1}^{n_p}w_t^ka_{jt}^k}.
\end{align}
Moreover, the MMSE estimate of $a_{jt}$ can be computed as
\begin{align}
\widehat{a}^{MMSE}_{jt}\coloneqq u\Biggl(\sum_{k=1}^{n_p}w_t^ka_{jt}^k-\frac{1}{2}\Biggr),
\end{align}
where $u$ denotes the step function.

\begin{figure}
\centering
\includegraphics[width=0.5\columnwidth]{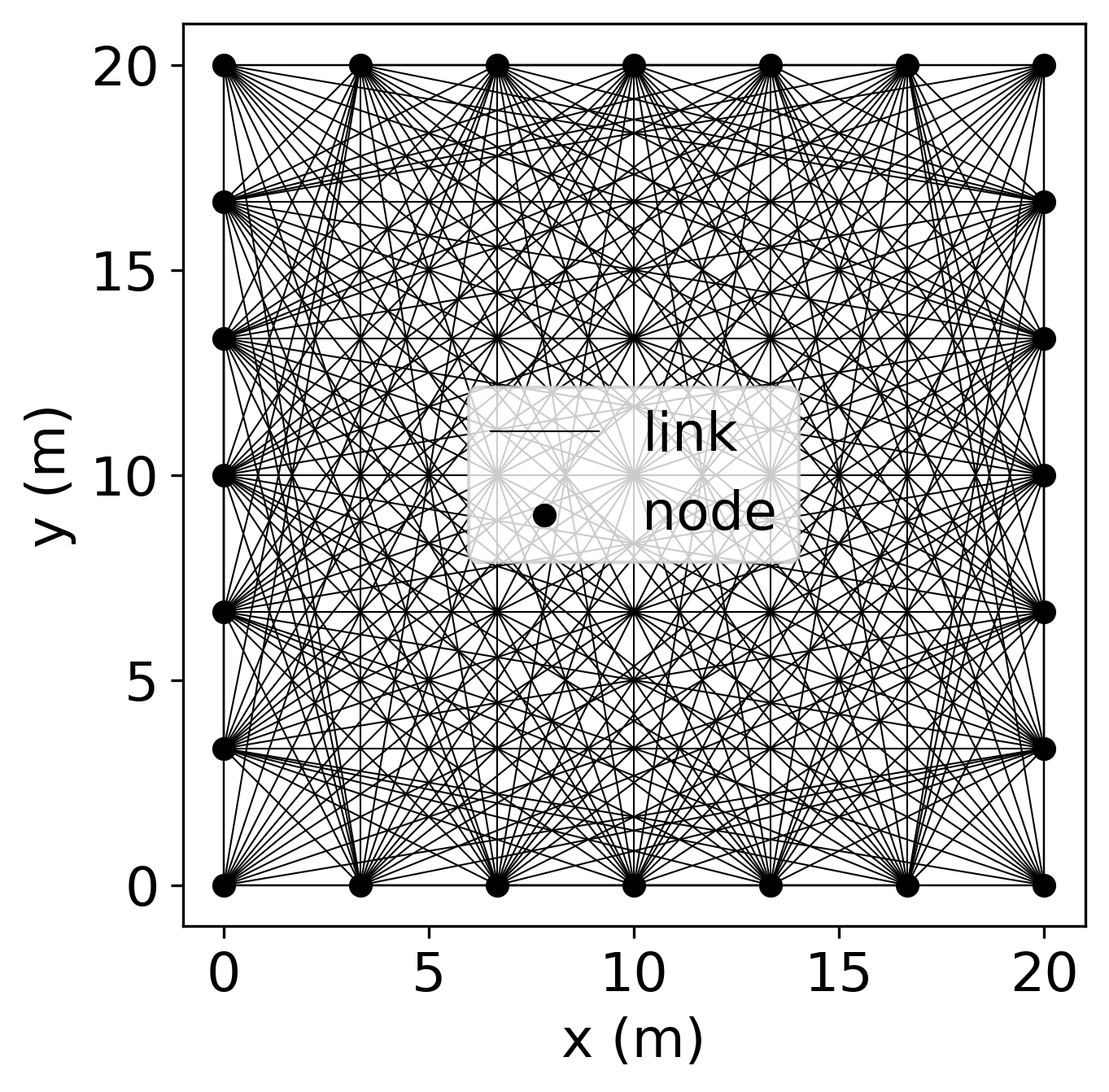}\vspace{-3mm}
\caption{Configuration of an RF antenna network.}
\label{fig:link}
\end{figure}
\begin{figure}
\centering
\includegraphics[width=0.55\columnwidth]{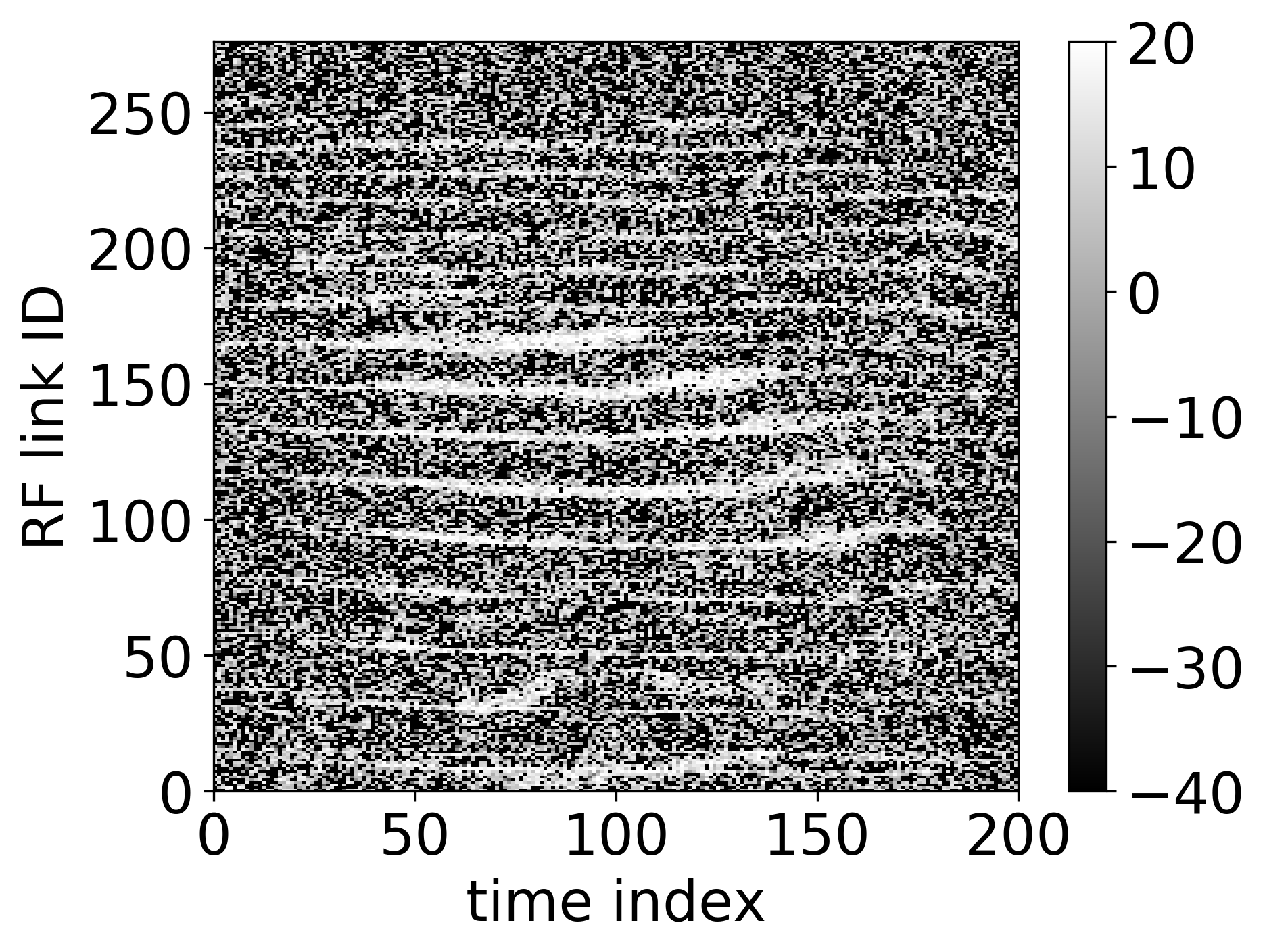}\vspace{-3mm}
\caption{Example of observed signals (SNR$=-5$\,dB).}\vspace{-2mm}
\label{fig:obs}
\end{figure}

\begin{figure}
\centering\vspace{-6mm}
\includegraphics[width=0.9\columnwidth]{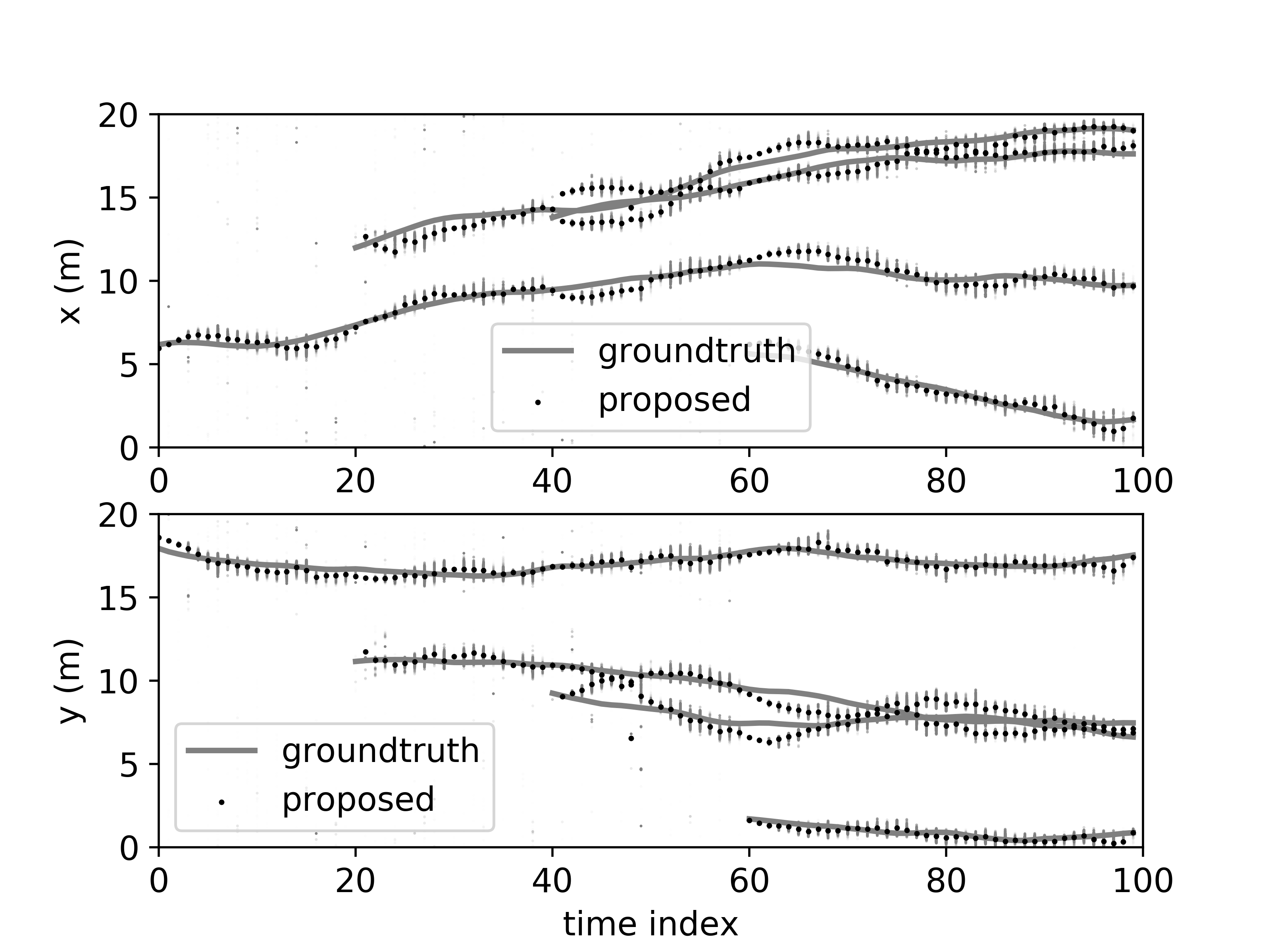}\vspace{-3mm}
\caption{Estimated x- and y-coordinates versus time; proposed method (SNR$=-5$\,dB).}
\label{fig:pos_prop}
\end{figure}

\begin{figure}
\centering\vspace{-3mm}
\includegraphics[width=0.9\columnwidth]{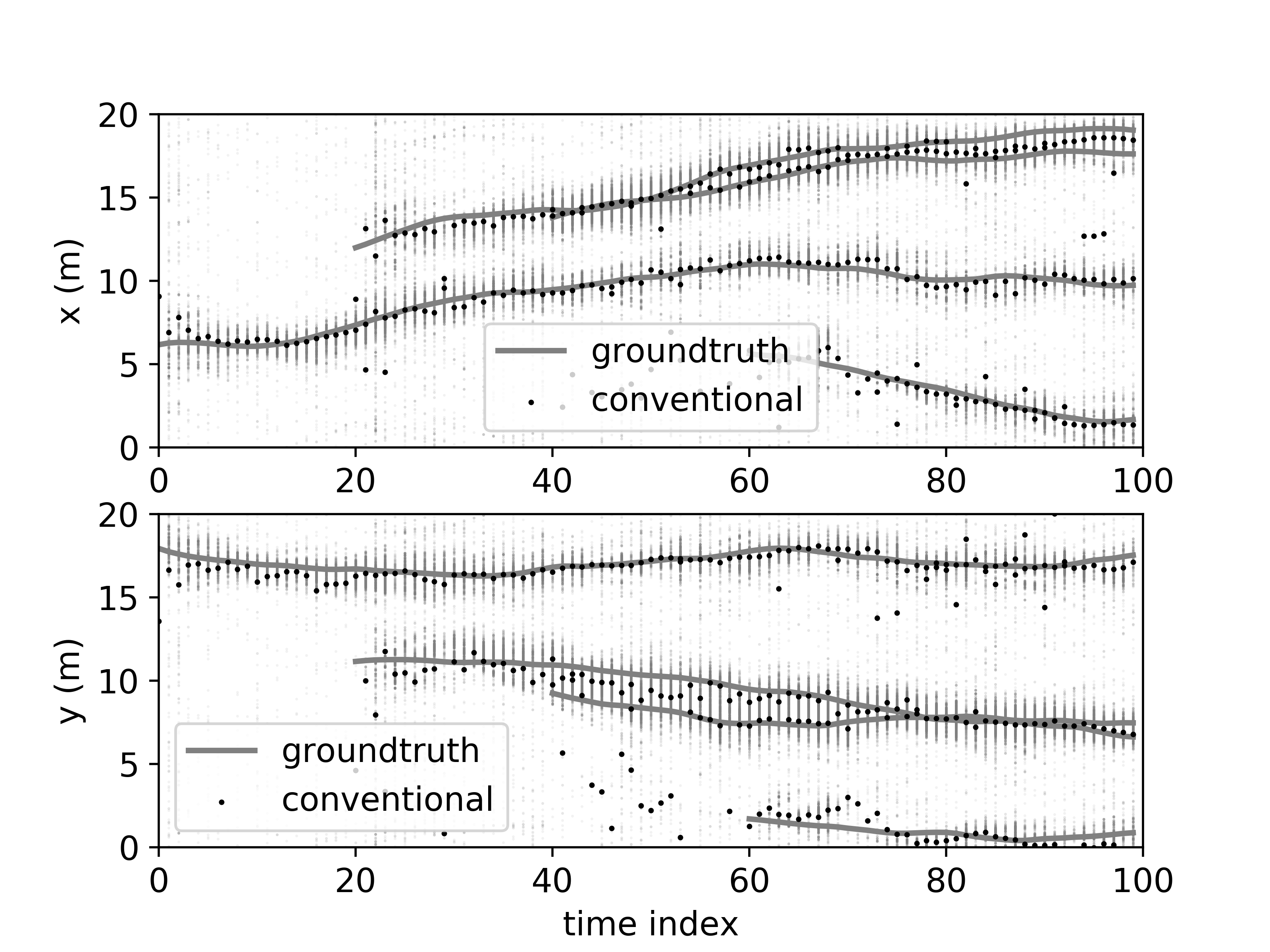}\vspace{-3mm}
\caption{Estimated x- and y-coordinates versus time; conventional method (SNR$=-5$\,dB).}
\label{fig:pos_conv}\vspace{-2mm}
\end{figure}

\section{Simulation: MTT for Radio-Frequency Tomography}
As an example, we considered a challenging task of MTT for an unknown, time-varying number of targets
with unknown initial positions in the context of radio-frequency (RF) tomography as in~\cite{Nannuru2013}.
RF tomography~\cite{Patwari2010,Wilson} aims to localize/track targets ({\it e.g.}, persons)
 in a surveillance region 
by using a network of RF antennas.
As in Fig.~\ref{fig:link}, we employed
 $n_a=24$ antennas (nodes)
on the perimeter of a square surveillance region 
of dimensions $20\,\text{m}\times 20\,\text{m}$.
In RF tomography,
signals communicated between RF antennas are used
instead of those emanating from targets.
These signals contain
target location information in the form of
 attenuation in received signal strength (RSS), 
which can be exploited for localization/tracking.
Hence, we used as sensor signals $\mathbf{z}_t$ RSS attenuation 
for all 
$
n_z=\frac{1}{2}n_a(n_a-1)=276$
antenna pairs (links).
On the other hand, states $\widetilde{\mathbf{x}}_t$ consisted
of kinematic states $\mathbf{x}_{jt}=(x_{jt},\dot{x}_{jt},y_{jt},\dot{y}_{jt})$
and an activity state $a_{jt}$ of each target $j$,
where $x_{jt}$ and $y_{jt}$ are Cartesian coordinates of target $j$ and 
$\dot{x}_{jt}$ and $\dot{y}_{jt}$ its velocities.

Transition of the states $\mathbf{x}_{jt}$ for a surviving target
was described by a linear Gaussian model
$
\mathbf{x}_{jt}=\mathbf{Fx}_{j,t-1}+\mathbf{Gw}_{j,t-1}
$~\cite{Nannuru2013}.
Here,
\begin{align}
&\mathbf{F}\coloneqq
\begin{pmatrix}
1&0\\
0&1
\end{pmatrix}\otimes
\begin{pmatrix}
1&T\\
0&1
\end{pmatrix},&\mathbf{G}\coloneqq
\begin{pmatrix}
1&0\\
0&1
\end{pmatrix}\otimes
\begin{pmatrix}
\frac{T^2}{2}\\
T
\end{pmatrix},
\end{align}
$\otimes$ is the Kronecker product,
$T=0.25\,\text{s}$ the sampling period,
and $\mathbf{w}_{j,t-1}$ zero-mean white Gaussian noise with covariance matrix $\bm{\Sigma}_{w}=0.35\,\mathbf{I}$.
The distribution $p_b$ for a newborn target was modeled by $p_b(x,\dot{x},y,\dot{y})=\mathcal{U}(x,y)\mathcal{N}(\dot{x}\mid 0,1)\mathcal{N}(\dot{y}\mid 0,1)$, where $\mathcal{U}$ denotes the uniform distribution over the surveillance region.
The transition probability matrix (\ref{eq:TPM}) for $\mathbf{a}_t$ was given by $\pi_b=0.2$ and $\pi_d=0.1$.
The initial state distribution was modeled by $p({\mathbf{x}}_0,\mathbf{a}_0)=\prod_{j=1}^{n_{max}}\{p(\mathbf{x}_{j0})P(a_{j0})\}$, 
where $p(\mathbf{x}_{j0})$ was defined in the same way as $p_b$ and $P(a_{j0}=0)=1$.
The observation distribution was given by the superpositional model in (\ref{eq:Gaussian})
with zero-mean white Gaussian noise with covariance matrix $\bm{\Sigma}_v=\sigma_v^2\mathbf{I}$. The nonlinear function $h=(h_1,\ldots,h_{n_z})$ was given by
$
h_i(\bm{\xi})=\phi\exp(-\frac{d_i(\bm{\xi})}{\sigma_h}),
$
where $d_i(\bm{\xi})$ is an elliptical distance~\cite{Chen2011}
between the $i$th link and a target with states $\bm{\xi}$ and $\phi=5$ and $\sigma_h=0.2$ are
 empirically determined hyperparameters. 

\begin{figure}
\centering
\includegraphics[width=0.6\columnwidth]{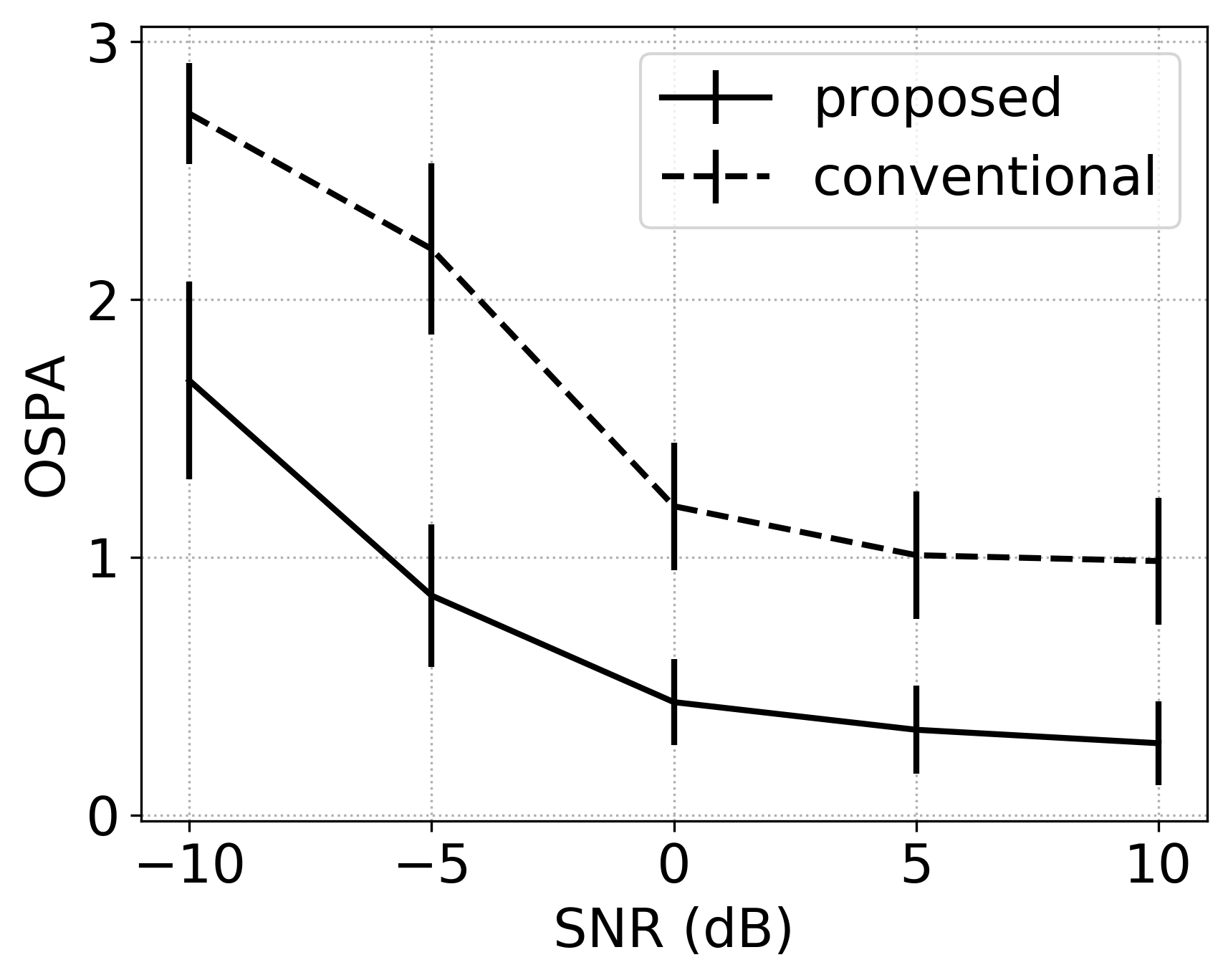}\vspace{-3mm}
\caption{OSPA metric versus SNR.}\vspace{-2mm}
\label{fig:OSPA}
\end{figure}

Sensor signals were generated as follows. 
Temporal behavior of $\mathbf{a}_t$ was
deterministically scheduled so that the number of active targets started from one, 
then increased gradually up to four, 
and finally decreased gradually down to one.
On the other hand, $\mathbf{x}_t$ was randomly generated based on the above model, and so was
$\mathbf{z}_t$. We adjusted
 $\sigma_v^2$ to give a desired signal-to-noise ratio (SNR), where $SNR\,(dB)\coloneqq10\log_{10}\langle\|\sum_{j=1}^{n_{max}}a_{jt}h(\mathbf{x}_{jt})
\|_2^2\rangle
-10\log_{10}\langle\|\mathbf{v}_t\|_2^2\rangle$ with $\langle\cdot\rangle$ being temporal averaging.
The number of time steps were 200,
corresponding to $50\,\text{s}$.
Figure~\ref{fig:obs} shows an example of  observed signals.

The proposed method was compared with
an
approximate $\Sigma$-CPHD filter~\cite{Nannuru2013},
which we hereafter call the conventional method.
The number of particles was fixed to $n_p=2000$ in the proposed method,
and time-varying with 
 500 particles per target plus 500 particles for proposing newborn targets in the conventional method.
In both methods, an auxiliary particle filter with
residual resampling~\cite{Liu1998} was employed.
The maximum possible number of targets was set to $n_{max}=4$ in the proposed method, and to $10$ in the conventional method.
In the conventional method, 
the probability of a target being born was set to 0.03, and
 the probability of each target surviving to 0.985. 

Figures \ref{fig:pos_prop} and \ref{fig:pos_conv} show estimated x- and y-coordinates versus time for the proposed and the conventional methods, respectively.
These estimates were obtained by MMSE estimation and $k$-means clustering in the proposed and the conventional methods, respectively.
Particles are also shown by gray dots with associated weights expressed by
darkness.
Figure~\ref{fig:OSPA} shows the estimation error in terms of optimal subpattern assignment (OSPA) metric~\cite{Schuhmacher2008} as a function of the SNR. The error bar shows (the mean) $\pm$ (one standard deviation) for 100 trials.

\section{Conclusion}
In this paper, we proposed
a particle filter for
multi-target track-before-detect using
superpositional sensor signals.
A simulation example of MTT for RF tomography clearly showed effectiveness of the proposed method.
Future work includes state augmentation with unknown signal amplitudes~\cite{Rollason2001,Rollason2018} and
estimation of static parameters ({\it e.g.}, $\pi_b$ and $\pi_d$).

\bibliographystyle{IEEEtran}
\bibliography{masterlist_v4b}
\end{document}